\documentclass[twocolumn]{revtex4}

\usepackage{graphicx}

\begin{document}

\title{ Sociophysics: A review of Galam models}

\author{Serge Galam}
\email{serge.galam@polytechnique.edu}
\affiliation{Centre de Recherche en \'Epist\'emologie Appliqu\'ee (CREA),\\
\'Ecole Polytechnique and CNRS, \\1 Rue Descartes, 75005 Paris, France}

\begin{abstract}
We review a series of models of sociophysics introduced by Galam and Galam et al in the last 25  years. The models are divided in five different classes, which deal respectively with democratic voting in bottom up hierarchical systems, decision making, fragmentation versus coalitions, terrorism and  opinion dynamics. For each class the connexion to the original physical model and technics are outlined underlining both the similarities and the differences. Emphasis is put on the numerous novel and counterintuitive results obtained with respect to the associated social and political framework. Using these models several major real political events were successfully predicted including the victory of the French extreme right party in the 2000 first round of French presidential elections, the voting at fifty - fifty in several democratic countries (Germany, Italy, Mexico), and the victory of the no to the 2005 French referendum on the European constitution. The perspectives and the challenges to make sociophysics a predictive solid field of science are discussed.
 \end{abstract}


\maketitle

\newpage

\section{Introduction}

The field of sociophysics has emerged in the 19-seventies with only a very few scarce papers in the hostile environment of the community of physicists. It started to attracted some  physicists around the mid nineties and then took over fifteen years to nucleate involving a growing number of physicists. Today it is a recognized  field of physics anchored within statistical physics. It is flourishing and expanding with hundreds of papers published in the leading physical journals and quite a few international conferences held each year.

The topics covered by sociophysics are becoming numerous and address many different problems including  social networks, language evolution, population dynamics, epidemic spreading, terrorism, voting, coalition formation and opinion dynamics. Among these topics the subject of opinion dynamics has become one of the main streams of sociophysics producing a great deal of  research papers also in this journal, including this issue. 

This review does not deal with all of these papers because of the restriction made clear by its title. This does not mean that the other papers are less important or worse than those cited here. But we restrict the presentation to the models introduced by Galam and Galam et al over the last twenty five years, a good part of them being the pioneer works of sociophysics. A Springer book is in preparation on the subject. These models deal  with the five  subjects of democratic voting in bottom up hierarchical systems, decision making, fragmentation versus coalitions, terrorism and  opinion dynamics. 

The first class of models \cite{v1, v2, v3, v4, v5, v6, v7, v8, v9, v10, v11, v12, v13} consider a population, which is a mixture of two species A and B. A bottom up hierarchy is then built from the population using local majority rules with the possibility of some power inertia bias. Tree like networks are thus constructed, which combine a random selection of agents at the bottom from the surrounding population with an associated deterministic outcome at the top. The scheme relates on adapting  real space renormalization group  technics to build a social and political structure. 

The second class \cite{s1,s2, s3, s4, s5, s6, s7, s8, s9, s10} tackles the problem of decision making in various frames including  firms and small committees. It uses ferromagnetic Ising spin Hamiltonians with both external and random quenched fields at both zero and non zero temperatures. The associated phase diagrams are constructed. The effect of reversing an external field on the collective equilibrium state is studied with an emphasis on the existence of nucleation phenomena. Mean field treatment is applied.

The third class \cite{f1, f2, f3, f4, f5} introduces a combination of random bond and random site spins glasses to describe the formation of coalitions as well the dynamics of fragmentation among a group of countrys. External and local fields are also considered together with site dilution effects in mixtures of  ferro and anti-ferromagnetic spin Hamiltonians. Ising and Potts variables are used.

The fourth class \cite{t1, t2, t3, t4, t5, t6} studies some aspects of terrorism by focusing on the role of passive supporters in the creation of the open social spaces, which are opened to terrorist activities. It relies on the theory of percolation and uses the dependence of the percolation threshold upon the space dimensionality. 

The fifth class \cite{o1, o2, o3, o4, o5, o6, o7, o8, o9, o10, o11, o12, o13,  o14, o15, o16} investigates opinion dynamics within reaction-diffusion like models. Two and three states variables are used. Three king of agents are also considered, which are respectively floaters, contrarians and inflexibles. The dynamics operates via local updates and reshuffling. Technics from real space renormalization group approach are used.

For each class of models the precise connexion to the original physical model is made. Similarities and differences are outlined emphasizing the eventual novelties with respect to the statistical physics counterparts. 

The numerous results obtained by each class of models are reviewed enlightening the novel and counterintuitive aspects with respect to the associated social and political framework. In particular several major real political events were successfully predicted using these models. It includes the victory of the French extreme right party in the 2000 first round of French presidential elections  \cite{vp1, vp2, vp3, vp4, vp5, vp6}, the voting at fifty - fifty in several democratic countries (Germany, Italy, Mexico) \cite{op6, op7, op8}, and the victory of the no to the 2005 French referendum on the European constitution \cite{op9}. 

To conclude, the perspectives to make sociophysics a predictive solid field of science are discussed, emphasizing both the challenges and the risks.


\section{Bottom-up voting in hierarchical systems}

The main question of this class of models is to measure the effective democratic balance of hierarchical organizations based on local bottom up voting using local majority rule. The net result is the singling out of a series of anti-democratic effects. In particular a model for a seemingly democratic dictatorship is derived \cite{v1, v2, v3, v4, v5, v6, v7, v8, v9, v10, v11, v12, v13}. The models shed a new counter intuitive light of some surprising major political events, which occurred in the recent years  \cite{vp1, vp2, vp3, vp4, vp5, vp6}.

The main scheme considers a population with a two species A and B mixture, whose respective proportions are respectively $p_0$ and  $(1-p_0)$.  It could be either a political group, a firm, or a whole society. At this stage each member does have an opinion. From now on we will use the political language. A bottom up hierarchy is then built by extracting randomly some agents from the surrounding population.  These agents are distributed randomly in a series of groups  with  finite size $r$, which constitute the hierarchy bottom. It is the level (0) of the hierarchy.

\begin{figure}
\includegraphics[width=\columnwidth]{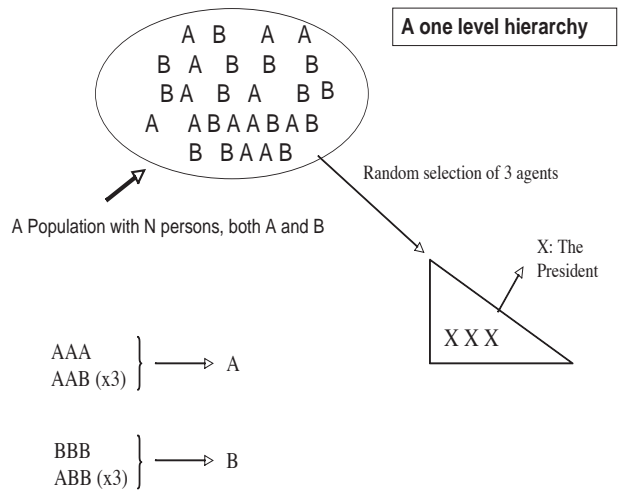}
\caption{A one level hierarchy with three agents randomly selected from a population. Then, they elect the president using a majority rule.
}
\label{n1}
\end{figure}

Then each one of these groups elects a representative according to some well defined voting rule $R_r(p_0)$, which is a majority rule function of the current composition of the group. This composition is probabilistic and depends on $p_0$ since the group members are randomly selected from the surrounding population. Therefore the voting outcome of the group is either an A with a probability
\begin{equation}
p_1=R_r(p_0) \
\end{equation}
or a B with probability $(1-p_1)$ as shown in Figure (\ref{n1}).

The ensemble of elected representatives constitute the first level of the hierarchy. The number of elected representatives is  a fraction of the number of bottom agents since $r$ agents elects one representative. Once they have been elected they formed another series of finite size groups, which in turn elected higher level representative according to the same voting rule used to elect them. The process is repeated  up forth with $p_n=R_r(p_{n-1})$ till one  upper level is constituted by one single group, which eventually elects the hierarchy president. One typical hierarchy is exhibited in Figure (\ref{n2}).

To exemplify the effects of such a bottom up voting scheme we study  two simple special cases, which embody all the main features of the process. All voting groups at each hierarchical level are set to the same unique size and that size is taken equal to respectively 3 and 4.

\subsection{The local majority rule model}

We consider groups, which are constituted by randomly aggregating 3 agents \cite{v1, v2, v4, v5, v6, v7,  v9, v10, v11, v12}. It yields the probability to
have an A elected at level $(n+1)$ from a group at level $n$,
\begin{equation}
p_{n+1}\equiv P_3(p_n)=p_n^3+3 p_n^2 (1-p_n) \ ,
\end{equation}
where $p_n$ is the proportion of elected A persons at level-n.

The voting function $P_3(p_n)$ has 3 fixed points
$p_d=0$, $p_{c,3}=1/2$ and $p_t=1$. First one corresponds
to the disappearance of the A. Last one $p_t$ represents the
dictatorship situation where only A are present. Both are stable.
In contrast $p_c$ is unstable. It determines the threshold to full
power. Starting from $p_0<1/2$ repeating voting leads
towards (0) while the flow is in direction of (1) for
$p_0>1/2$.

Therefore majority rule voting produces the self-elimination of
any proportion of the A-tendency as long as $p_0<1/2$,
provided there exists a sufficient number of voting levels. It is therefore essential to determine the number of levels required to ensure full leadership to the initial larger
tendency. 

For instance starting from $p_0=0.43$ we get successively
$p_1=0.40 $, $p_2=0.35$, $p_3=0.28$, $p_4=0.20$, $p_5=0.10$,
$p_6=0.03$ down to $p_7=0.00$. Therefore 7 levels are sufficient
to self-eliminate $43\%$ of the population.

Though the aggregating voting process eliminates a tendency it stays
democratic since it is the leading tendency (more than $50\%$),
which eventually gets the total leadership of the organization. The situation
is symmetry with respect to A and B. Many countries apply the corresponding winner-takes-all rule, which gives power to the winner of an election.

\begin{figure}
\includegraphics[width=\columnwidth]{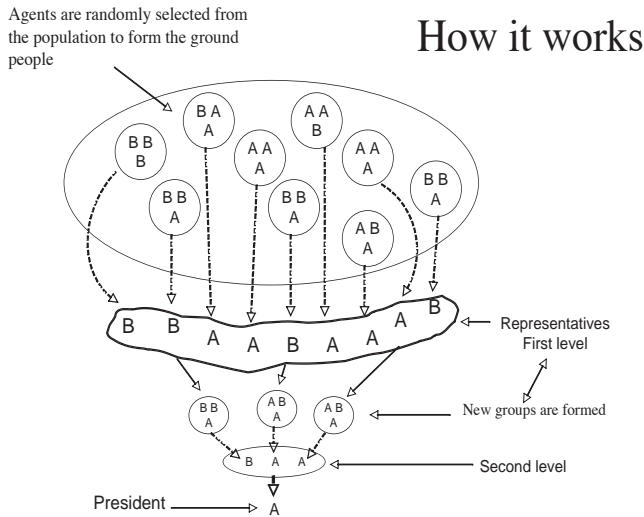}
\caption{A three level hierarchy with groups of 3 persons.
}
\label{n2}
\end{figure}

\begin{figure}
\includegraphics[width=\columnwidth]{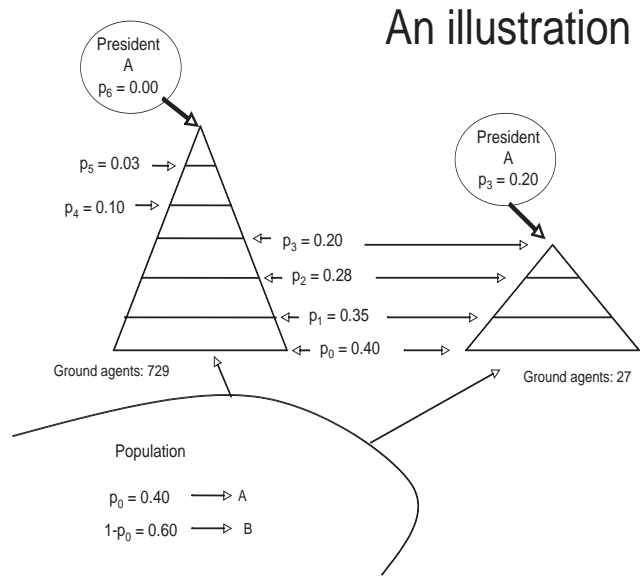}
\caption{Two different three agent voting groups hierarchies built from the same population where A has a support of $p_0=0.40$ and B a support of $0.60$. In the left side hierarchy, six levels restore the democratic balance with $p_1=0.00$. It involves 729 agents at the bottom with a total of 1093 agents. The right side hierarchy has only three levels making the president an A with a probability $p_3=0.20$. The bottom involves 27 agents for a total of 40 agents. }
\label{n3}
\end{figure}

The series of Figures (\ref{n1}, \ref{n2}, \ref{n3}) illustrate the scheme of building a bottom up democratic hierarchies. The first one is the simplest with one level, the presidential one. The second one shows a $n=3$ hierarchy while the last one increase the number of levels to $n=6$. The three of them can be built from the same surrounding population.

\subsection{Including  power inertia into the local majority rule}

We now constitute groups of 4
persons instead of 3. It yields the 2A-2B
configurations for which there exists no majority.
The question is then how to solve such a case.

In most
social situations it is well admitted that to change a policy
required a clear cut majority. In case of no decision, things
stay as they are. It is a natural bias in favor of the status quo \cite{v1, v2, v4, v5, v6, v7,  v9, v10, v11, v12}.
In real institutions, such a bias is possibly avoided,
for instance giving one additional
vote to the committee president.

Accordingly the voting function becomes non symmetrical.
Assuming the B were in power, for an A to be elected at level
$n+1$ the probability becomes,
\begin{equation}
p_{n+1}\equiv P_4(p_n)=p_n^4+4 p_n^3 (1-p_n) \ ,
\end{equation}
while for B it is,
\begin{equation}
1- P_4(p_n)=p_n^4+4 p_n^3 (1-p_n)+2 p_n^2 (1-p_n)^2 \ ,
\end{equation}
where last term embodies the bias in favor of B. The associated stable
fixed points are unchanged at  (0) and (1). However the unstable one is drastically shifted to,
\begin{equation}
p_{c,4}=\frac{1+\sqrt{13}}{6} \approx 0.77\ ,
\end{equation}
which sets the threshold to power for A at a much higher value than the expected fifty percent. In addition, the process of self-elimination is accelerated. For instance starting from
$p_0=0.69 $ yields the series  $p_1=0.63 $, $p_2=0.53 $,
$p_3=0.36 $, $p_4=0.14 $, $p_5=0.01 $, and $p_6=0.00$. The series shows how $63\%$ of a
population disappears democratically from the leadership levels within only 5 voting
levels.

Using an a priori reasonable bias in favor of the B turns a
majority rule democratic voting to an effective dictatorship outcome. Indeed
to get to power the A must pass over $77\%$ of support, which is
almost out of reach in any democratic environment.


\subsection{Larger voting groups and the magic formula}

For voting groups of any size $r$ the voting
function $p_{n+1}=P_r(p_{n})$ writes,
\begin{equation}
P_r(p_n)=\sum_{l=r}^{l=m}\frac{r!}{l!(r-l)!} p_n^l(1+p_n)^{r-l}\ ,
\end{equation}
where $m=(r+1)/2$ for odd $r$ and $m=(r+1)/2$ for even $r$ to
account for the B favored bias. The two stable fixed points $p_d=0$ and $p_t=1$ are independent of the group size $r$.
The unstable $p_{c,r}=1/2$ is also independent of the group size $r$ for odd values of $r$, for which there exist no bias. On the contrary it does vary with $r$ for even values. It starts at $p_{c,2}=1$ for $r=2$, decreases to $p_{c,4}=(1+\sqrt{13})/6 \approx 0.77$ for $r=4$ and then keeps on decreasing asymptotically towards $1/2$ from above \cite{v1, v2, v4, v5, v6, v7,  v9, v10, v11, v12}.

When $p_0<p_{c,r}$ we can calculate analytically the critical number of levels $n_c$ at which $p_{n_c}=\epsilon$ with
$\epsilon$ being a very small number. This number determines the level of confidence
for the prediction to have no A elected from level $n$ and higher, i.e., only B elected. To make the evaluation we first expand the voting function $p_n=P_r(p_{n-1})$ around the unstable fixed point $p_{c,r}$,
\begin{equation}
p_n\approx p_{c,r}+(p_{n-1}-p_{c,r}) \lambda_r \ ,
\end{equation}
where $\lambda_r \equiv dP_r(p_n)/dp_n|_{p_{c,r}}$ with
$P_r(p_c)=p_{c,r}$. It can rewritten as,
\begin{equation}
p_n-p_{c,r}\approx (p_{n-1}-p_{c,r}) \lambda_r \ ,
\end{equation}
which then can be iterated to get,
\begin{equation}
p_n-p_{c,r}\approx (p_0-p_{c,r}) \lambda_r^n \ .
\label{v1}
\end{equation}
The critical number of levels $n_c$ at which $p_n=\epsilon$ is then extracted by
taking the logarithm on both sides to obtain,
\begin{equation}
n_c\approx -\frac{\ln (p_c-p_0)}{\ln \lambda_r} +n_0   \ ,
\end{equation}
where $n_0\equiv \ln (p_{c,r} -\epsilon)/\ln \lambda_r$. 
Putting in $n_0=1$ while taking the integer part of the expression yields rather good estimates of $n_c$ with respect to the exact estimates obtained by iterations. 

Above expression is interesting but does not allow to a define strategy from either A or B since most organizations have a fixed structure, which cannot be modified at will before every new election, even if it is done sometimes. The number of hierarchical levels is thus fixed and constant. Therefore to make the analysis useful the question  of
\begin{itemize}
\item How many levels are needed to get a tendency self eliminated?
\end{itemize}

becomes instead

\begin{itemize}
\item Given n levels what is the necessary overall support in the surrounding population to get full power with certainty?
\end{itemize}

Keep in mind that situations for respectively  A and B tendencies
are not always symmetric. In particular they are not symmetric for even size groups. Here we address the dynamics of voting with respect to the A perspective. To implement the reformulated  operative question, we rewrite Eq. (\ref{v1}) as,
\begin{equation}
p_0=p_{c,r}+(p_n-p_{c,r}) \lambda_r^{-n} \ ,
\label{v2}
\end{equation}
from which two different critical thresholds are obtained.
The first one is the disappearance threshold $p_{d,r}^n$ which gives the value
of support under which the A disappears with certainty from elected representatives  at level $n$, which is the president level. In other words, the elected president is a B with certainty. It is obtained putting $p_n=0$ in Eq. (\ref{v2}) with,
\begin{equation}
p_{d,r}^n=p_{c,r}(1-\lambda_r^{-n}) \ .
\end{equation}
In parallel putting $p_n=1$ again in Eq. (\ref{v2}) gives the second threshold $p_{f,r}^n$ above which 
the A get full and total power at the presidential level. Using Eq.(11), we get,
\begin{equation}
p_{f,r}^n=p_{d,r}^n+\lambda_r^{-n}\ .
\label{v3}
\end{equation}

The existence of the two thresholds $p_{d,r}$ and $p_{f,r}$ produces a new region $p_{d,r}^n<p_0<p_{f,r}^n$ in which the A neither disappears totally nor get full power with certainty. There $p_n$ is neither $0$ nor $1$). It is
therefore a region where some democraty principle is prevailing since results
of an election process are still probabilistic. No tendency is sure of winning
making alternating leadership a reality. 

Its extension is given by $\lambda_r^{-n}$ as from Eq. (\ref{v3}). It shows that the probability region shrinks as a power law of  the number $n$ of hierarchical levels.
Having a small number of levels puts higher the threshold to a total 
reversal of power but simultaneously lowers the threshold for democratic disappearance.

To get a practical feeling from Eq. (\ref{v3}) we look at the case $r=4$ where we have $\lambda =1.64$ and $p_{c,4}\approx 0.77$.
Considering $n=3, 4, 5, 6, 7$ level organizations, $p_{d,r}^n$ equals to 
respectively $0.59$, $0.66$, $0.70$, $0.73$ and $0.74$. In parallel $p_{f,r}^n$
equals $0.82$, $0.80$, $0.79$, $0.78$ and $0.78$. The associated range extension is 0.23, 0.14, 0.09, 0.05, 0.04. These series emphasizes drastically the dictatorship character of the bottom up voting process.

\subsection{Visualizing the dynamics: a simulation}

\begin{figure}
\includegraphics[width=\columnwidth]{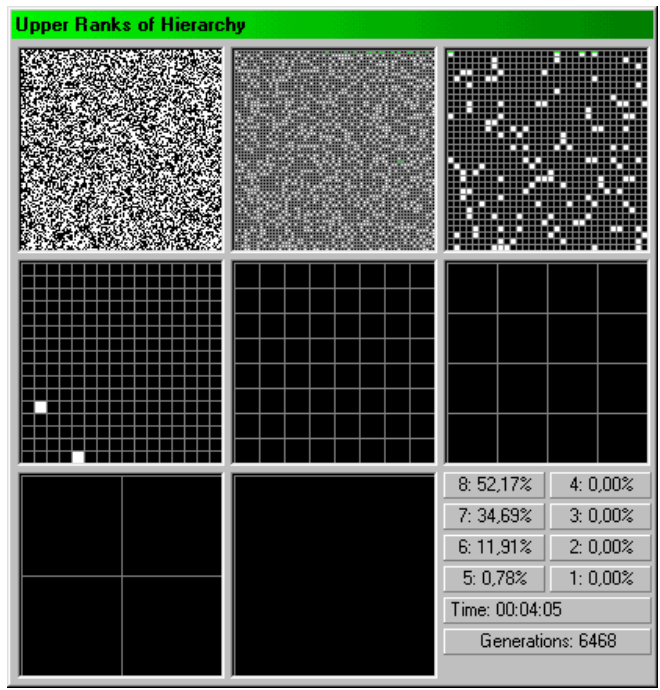}
\caption{A 8 level hierarchy for even groups of 4 persons.
The two A and B parties are represented respectively in white and black
with the bias in favor of the black squares, i. e., a tie 2-2 votes for a black
square. Written percentages are for the white
representation at each level. The ``Time" and ``Generations"
indicators should be discarded. The initial white support is $52.17\%$.
}
\label{s1}
\end{figure}

\begin{figure}
\includegraphics[width=\columnwidth]{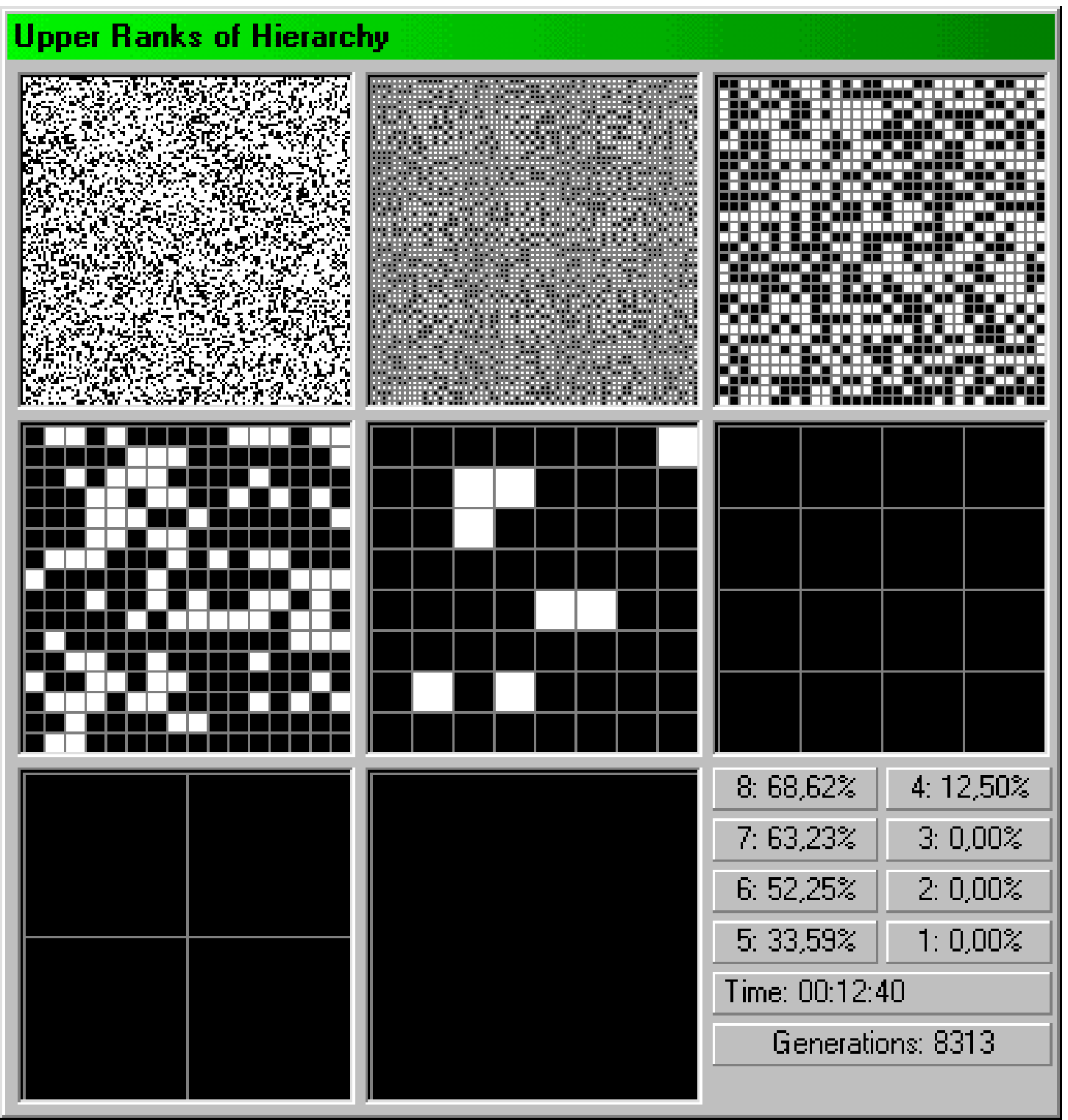}
\caption{The same as Figure 6 with an initial white support of
$68.62\%$. The presidency stays black.
}
\label{s2}
\end{figure}

\begin{figure}
\includegraphics[width=\columnwidth]{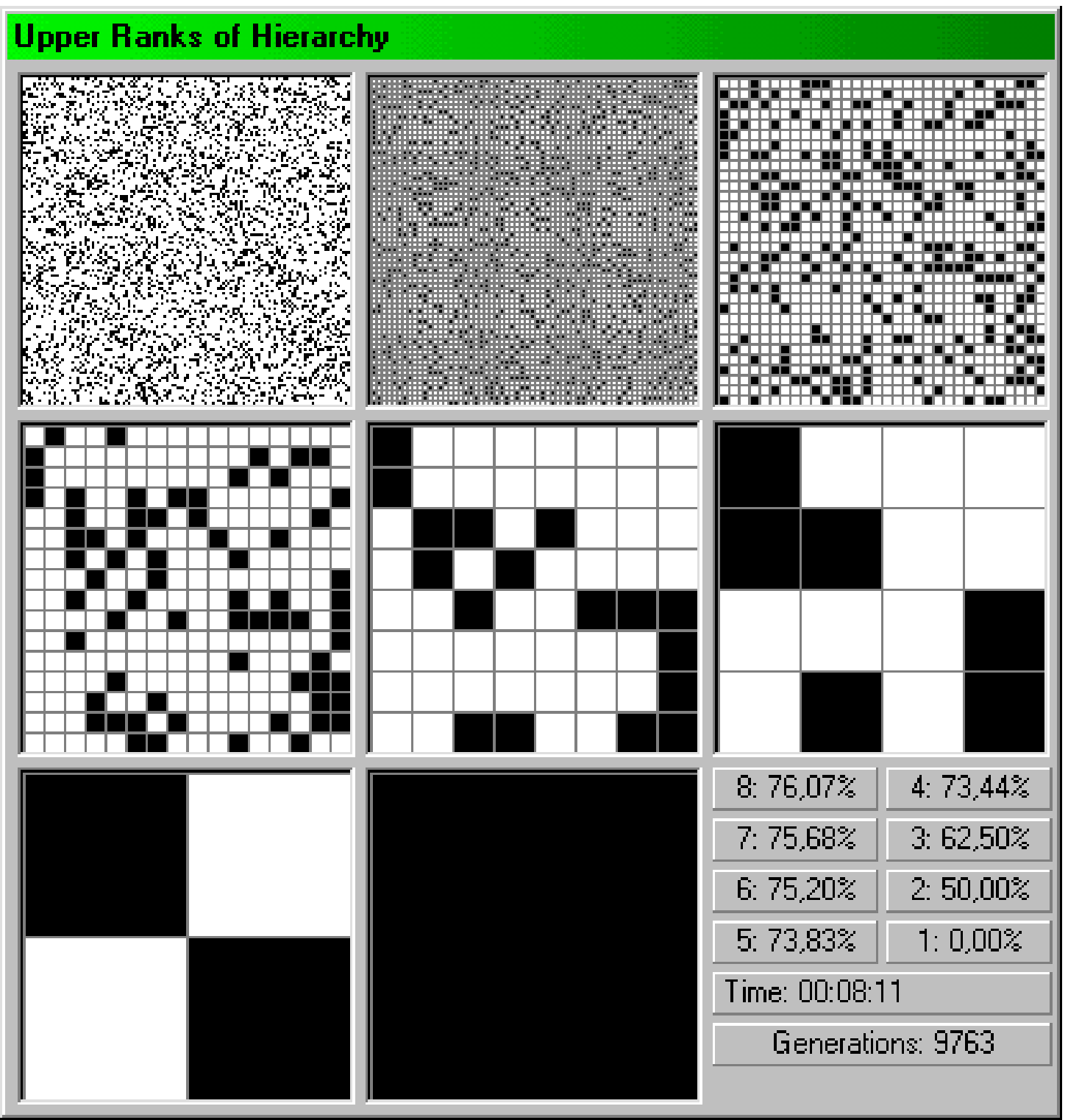}
\caption{The same as Figure 6 with an initial white support of
$76.07\%$. The presidency stays black.
}
\label{s3}
\end{figure}

\begin{figure}
\includegraphics[width=\columnwidth]{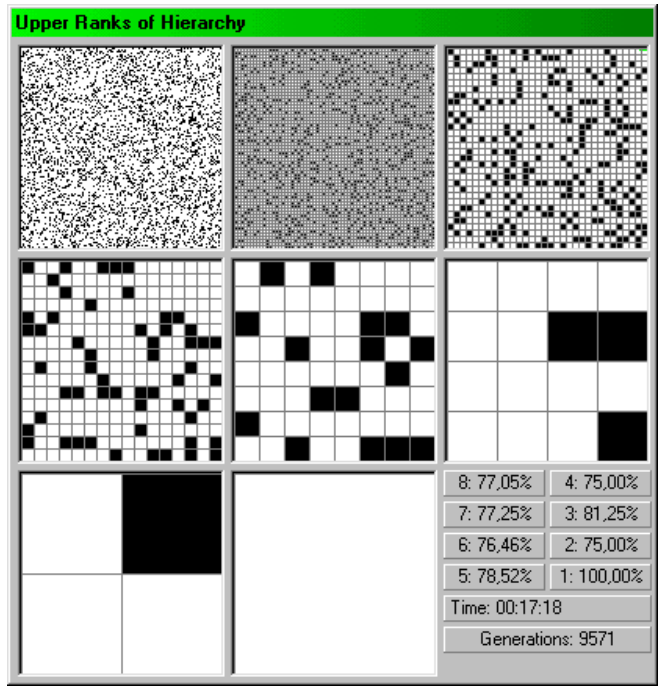}
\caption{The same as Figure 6 with an initial white support of
$77.05\%$. The presidency finally turned white.
}
\label{s4}
\end{figure}

To exhibit the phenomena a series of snapshots from a 
numerical simulation done with Wonczak \cite{v8} with 16384 agents are shown in Figures  (\ref{s1}, \ref{s2}, \ref{s3}, \ref{s4}). The two A and B parties are represented
respectively in white and black squares with the bias in favor of the blacks. The bottom up hierarchy operates with voting groups of size 4 and has 8 levels including the hierarchy bottom.

Four different initial bottom proportions of A and B are shown. On the first three pictures, a huge bottom white square majority is seen to get self-eliminated rather quickly. Written percentages on the lower right part are for the white representation at each level denoted (8) for the bottom and (1) for the president. The ``Time" and ``Generations"
indicators should be discarded.

Figure (\ref{s1}) shows $52.17\%$ bottom people in support of the   A (white sites in figures), a bit over the expected $50\%$ democratic threshold to take over. However,  3 levels higher no white square appears. The bottom majority has self-evaporated.

Figure (\ref{s2}) shows the same population with now a substantial increase in  A (white sites in figures) support with a majority of $68.62\%$, rather more than the democratic balance of $50\%$. And yet, after 4 levels no more white (A) square is found.

The situation has worsened in Figure (\ref{s3}) where the  A (white sites in figures) support has climbed up to the huge value of $76.07\%$. But again, 7 levels higher a B (black) is elected with certainty though its bottom support is as low as (black) $23.03\%$.

Finally Figure (\ref{s4}) shows an additional  very small $0.08\%$  increase in  A (white sites in figures) support, 
putting the actual support at  $77.05\%$, which in turn prompts a A to get elected (white site in Figures) president. 

These simulations provide some insight about  the often observed blindness of top leaderships towards huge and drastic increase of dissatisfaction at the botton level of an organization. Indeed it is seen how and why a president, who would get some information about the possible disagreement with its policy, cannot recognize the real current state for support in the population.  As seen from Figure, while the opposition is at already a height of $68.62\%$, the president gets $100\%$ of totally satisfied votes from the two levels below it. Accordingly it will conclude  at an overwhelming satisfaction, so why to make any policy change?

\subsection{Extension to 3 competing parties}

Up to now we have treated very simple cases to single out main trends produced 
by democratic voting aggregating over several levels. In particular we have 
shown how these thresholds become
non symmetric. Such asymetries are indeed always present in most realistic situations in particular when more than
two groups are competing.

 Let us consider for instance the case of three competing
groups A, B and C \cite{v3}. Assuming a 3-cell case, now the (A B C) configuration is unsolved
using majority rule as it was for the precedent (A A B B) configuration.
For the A B case we made the bias in favor of the group already in power, like
giving an additional vote to the comitee president. 

For multi-group competitions the situation is different. Typically the bias results from parties agreement.
For instance, in most cases  the two largest parties, say A and B are hostile one to the other while 
the smallest one C could compromise with either one of them. Then the (A B C) configuration gives a C elected. In such a case, we need 2 A or 2 B to elect respectively an A or a B. Otherwise a C is elected. Therefore the elective function for A and B are the same as for the AB $r=3$ model. It means that the critical threshold to full power to A and B is $50\%$. Acordingly, for initial A and B supports, which are lower than $50\%$ the C gets full power provided the number of levels is larger than some  minimum limit \cite{v3}. 

Generalization is possible to as many groups as wanted. However the analysis becomes very quickly much more heavy and must solve numercially. But the mean features of voting flows towards fixed point are preserved.

\subsection{Similarities and differences with the physical systems}

The model used does not have a direct statistical physics counterpart. Nevertheless it borrows from it two different features. The first one is to consider a mixture with two species A and B at fixed densities, but such a situation is not specific to physical models. Moreover it is worth to emphasize that although we are dealing with two species A and B, our agents are not Ising like variables. Each agent belongs to one party and does not change its affiliation. We are using a mixture of one state variables.

The second borrowed feature is the mathematical local bare mechanism of real space renormalization group scheme, which uses a majority rule to define a super spin. But the analogy ends there since our implementation is performed in a totally different way than in physics. 

In our voting case, the local rule is operated to add a real new agent above the given voting group. This agent does not substitutes to the group. It is not a virtual super spin. The group and its elected representative are real agents, which are simultaneously present. All the hierarchy levels are real, in the sense of the model.

In physics  renormalization group technics are a mathematical method.  It is used to extract rather accurately some physical quantities, which characterize a given system. Applying a renormalization group scheme to a physical system does not modify the system. In our case the system is built step by step following the scheme. The hierarchy does exist, it is the system. Accordingly a $n$ level hierarchy is different from a $m$ level hierarchy.

Given some proportions of respectively A and B agents in a population we build a bottom up voting hierarchy by selecting agents from that population. At the the bottom level of the hierarchy agents are randomly selected from the surrounding population to constitute the voting groups. Afterwards at the first level, the elected representatives are selected from the population according to the vote outcomes, i.e., the party affiliation is imposed by the voting of the group beneath. They are not randomly selected.

\subsection{Novel counterintuitive social and political predictions}

Although the model is only a snapshot of real hierarchies it graps some essential
and surprising mechanisms of majority rule voting. It is very generic and allows to consider many different applications.

In particular it exhibits several counterintuitive results and provides paradoxical and unexpected explanations to a series of social features and historical events. Especially the empirical difficulty in changing leaderships in well established institutions. It also allows to shed a new light on an astonishing and crucial historical event of the last century, the sudden and quick auto-collapse of eastern European communist parties.

Up to this historical and drastic end, communist parties in power has seemed to be eternal. Once they collapsed all once many explanations were given to base the phenomena on some hierarchical opportunistic change within the
various organizations.  Among others one reason for the eastern European countries was the end of the Soviet army threat. 

However our hierachical model may provide some different new insight at such a unique event. Communist organizations are indeed based, at least in principle,
on the concept of democratic centralism which is a tree-like hierarchy similar to our bottom up model. 
Suppose for instance that the critical threshold to power was of the order of $77\%$ like
in our size 4 case. We could then consider that the internal opposition to the orthodox leadership
did grow continuously over several decades to eventually turn massive, yet without any visible change at the top organizations. Then at some point of internal opposition, its internal increase has reached the critical threshold. There a little more increase would at once produce a surprising shift of the top leadership as exhibited in our series of Figures \cite{s1, s2, s3, s4}. From outside, the decades long increase of opposition was invisible. Indeed it looks like nothing was changing. And once the threshold passed, the shift appears as instantaneous \cite{vp3}.

Therefore, what looked  like a sudden and punctual  decision of a top leadership could be indeed the result of a very long and solid phenomenon inside the communist parties.
Such an explanation does not oppose to the very many additional
features which were instrumental in these collapses. It only  singles out some trend within 
the internal mechanism of these organizations, which in turn made them extremely stable.

Using our model we predicted a political scenario, which could happen in France, and which eventually did occur with respect to the extreme right party National Front. We enumerate the conditions for its success in 1997 \cite{vp1, vp2} and it happened along these line in 2000 with its leader winning the presidential first round. He eventually lost in the second final run \cite{vp4, vp5, vp6}.

\section{Group decision making}

Every person studying the Ising ferromagnetic model within the frame of modern statistical physics would envision an analogy with some social systems. It is a very appealing universal model, which could apply to a large spectrum of social situations. 

\subsection{The strike phenomena}

With Shapir and Gefen \cite{s1} we implemented the idea of using an Ising ferromagnetic system to describe the collective state of an assembly of agents, who can be in either one of two distinct individual states, to work or to strike, by choosing to study the collective phenomena of strike in firms. The ferromagnetic coupling being motivated by the social fact that people have the tendency to reproduce their neighbors attitude in particular in a conflicting situation.

A spin $\tilde{\mu_i}$ is associated to each agent $i$ with $i=1, 2, ..., N$. When  $\tilde{\mu_i}=0$ the agent is not working while $\tilde{\mu_i}=1$ means it is working at maximum individual production. A normalized individual production is then defined with
$\mu_i=2(\tilde{\mu_i}-1/2)$ giving an Ising variable $\mu_i=1$ if the agent is working and $\mu_i=-1$ when it is not working.

Two agents $i$ and $j$ interact via a coupling $J_{i,j}$. Depending on their synchronization, their respective behaviors create some dissatisfaction $-J_{i,j}\mu_i\mu_j$. To account for the social fact that people have the tendency to reproduce their neighbors attitude  the coupling is taking positive with $J_{i,j}>0$.

An external field $H=W-E$ is also applied, which couples linearly with each agent to create a dissatisfaction $-H\mu_i$ where $W$ is the actual wage and $E$ is the agent minimum salary expectation. When $W>E$ the salary is an incentive to work while $W<E$, the actual salary is perceived as not worth to work. 

Summing up above both contributions for all agents yields  the collective dissatisfaction function, which is indeed an Ising like Hamiltonian \cite{s1},
\begin{equation}
{\cal H}=-\sum_{(i,j)} J_{i,j}\mu_i\mu_j-H\sum_{i=1}^N \mu_i  ,
\label{strike}
\end{equation}
where $(i,j)$ denotes all interacting pairs of agents.

Introducing a social permeability $1/T$ where $T$ is the equivalent of the physical temperature, a global dissatisfaction function $F$, indeed a free energy, can be calculated. It is a function of the parameters $(M, K, H)$ where $M\equiv 1/N\sum_{i=1} ^N \mu_i$, $J_{i,j}=J$, $K\equiv T/J$ and Boltzmann's constant is set to unity.

\begin{figure}
\includegraphics[width=\columnwidth]{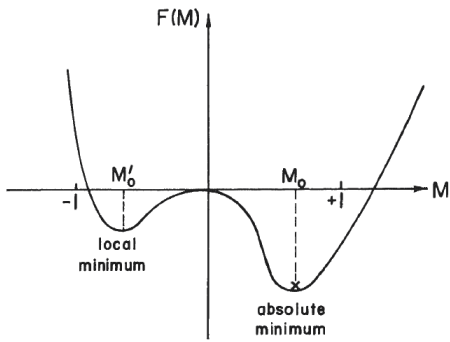}
\caption{Dissatisfaction function $F$ versus $M$ for $K<K_c$. the symbol x denotes the actual sate of the system.
}
\label{h1}
\end{figure}

\begin{figure}
\includegraphics[width=\columnwidth]{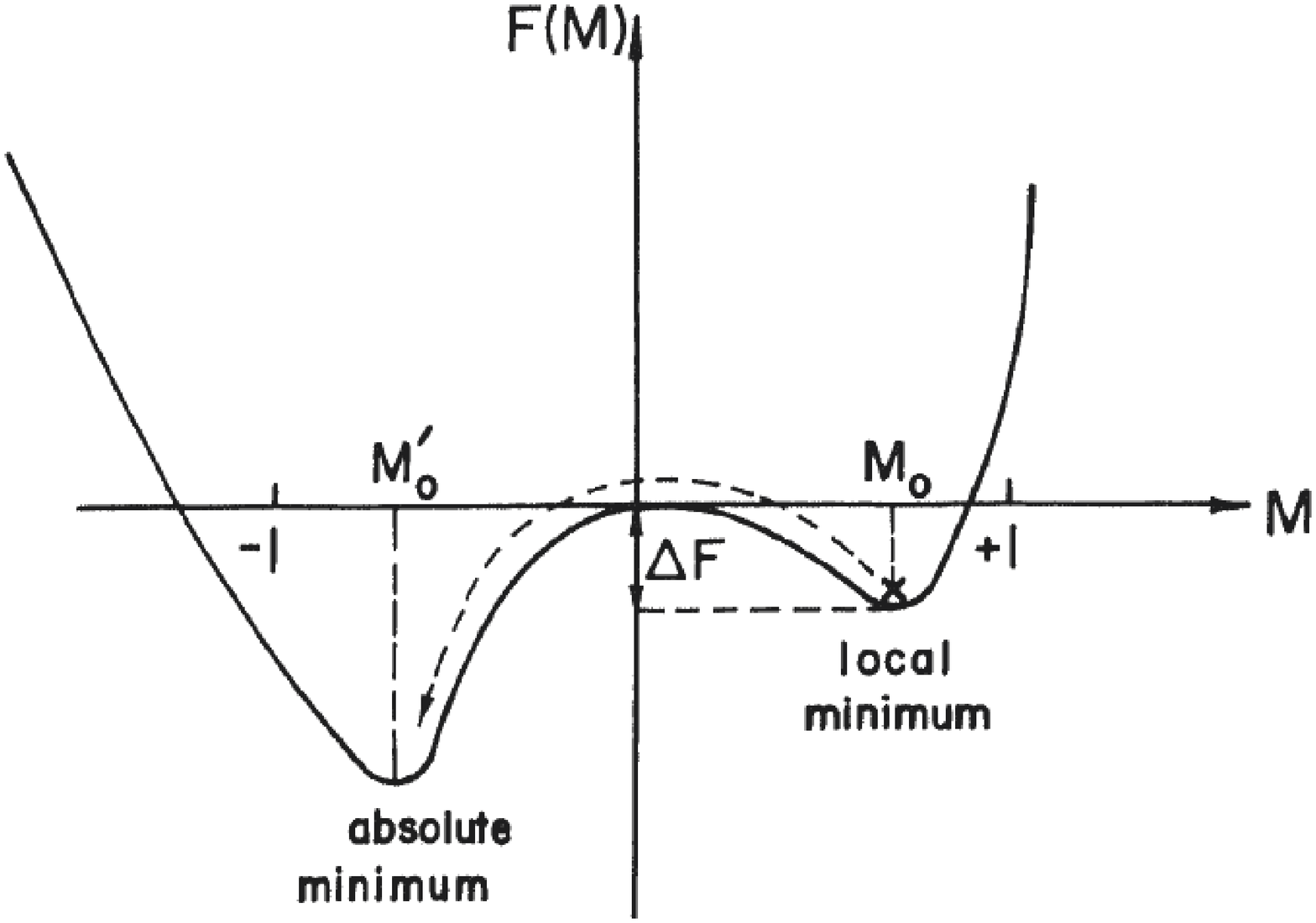}
\caption{A metastable case with $H<0$. The dash arrow indicates the eventual jump into a stable strike state driven by some external action.
}
\label{h2}
\end{figure}

\begin{figure}
\includegraphics[width=\columnwidth]{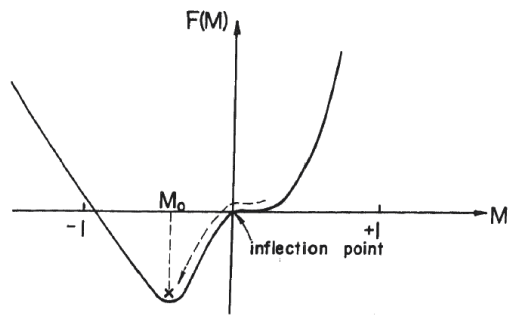}
\caption{The limit of metastability for $H<0$. $F$ has only one minimum.
}
\label{h3}
\end{figure}

\begin{figure}
\includegraphics[width=\columnwidth]{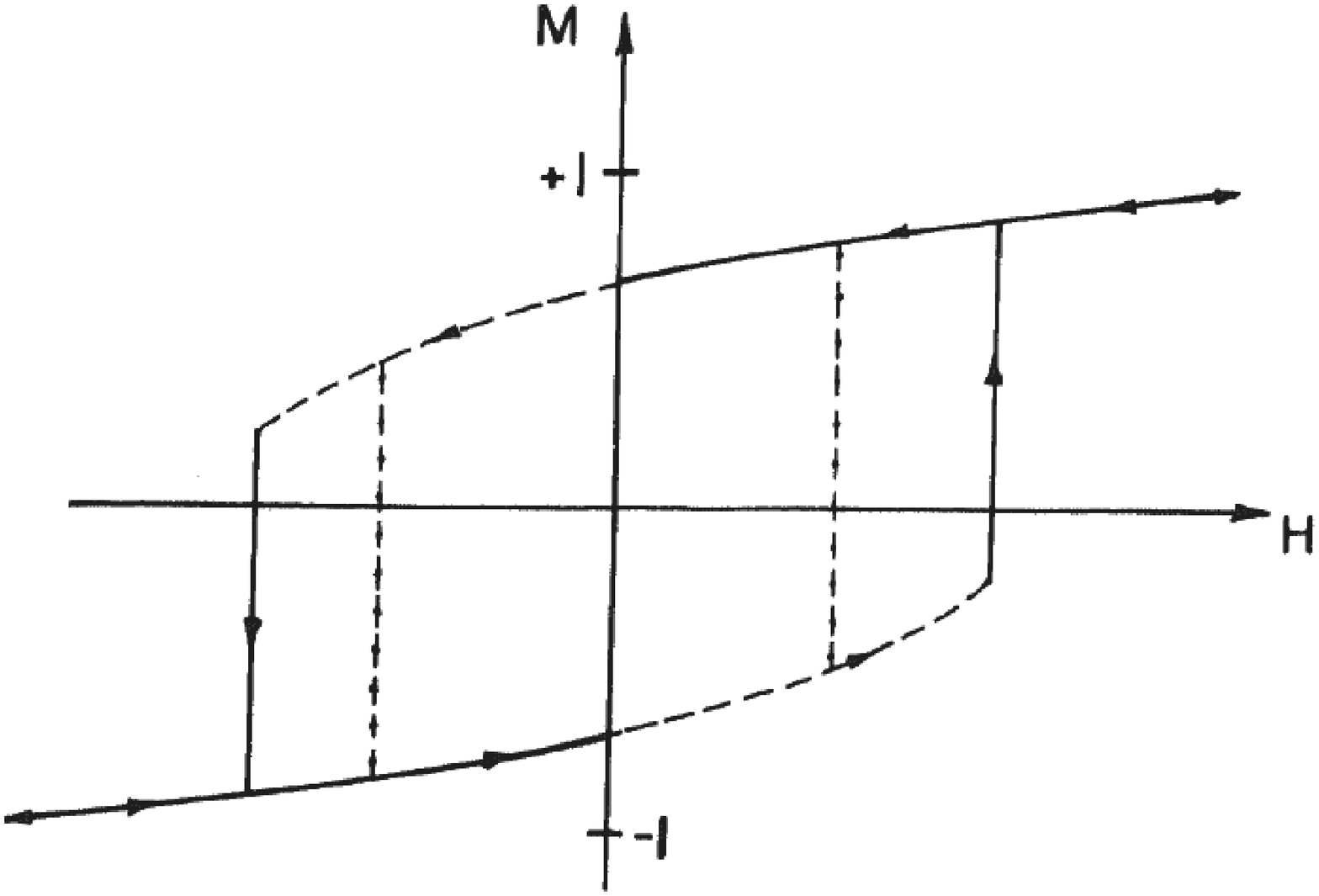}
\caption{$M$ versus $H$ for $K<K_c$. Arrows show the direction of change for $H$. Dashed lines denote metastable states. Dot-dash lines show possible jumps from metastable to stable states.
}
\label{h4}
\end{figure}

A principle of minimum dissatisfaction is postulated to determine the eventual equilibrium global states of the system of the assembly of $N$ agents. Performing a mean field treatment, all features of the ferromagnetic Ising model in an external field are recovered.
In particular the existence of two symmetric ordered phases with respectively $0<M\leq 1$ and $-1\leq M<0$ when $K<K_c$ where $K_c$ is a constant related to the coordination number. A phase transition is then found to occur at $K=K_c$ into a disordered phase with $M=0$. It is stable in the range $K>K_c$. Metastability and the nucleation phenomenon are also obtained by reversing the external field $H$ sign. The series of Figures (\ref{h1}, \ref{h2}, \ref{h3}, \ref{h4}).

\subsubsection{Similarities with physical systems and novel counterintuitive social results}

Here the model is identical to the Ising ferromagnet in an external field and the treatment is a classical mean field theory. All the rich associated properties are thus recovered. 
The difference dwells in the different frame in which the model is used. The novelty relies on the transfer from statistical physics to the new field of social sciences.  In contrast the import of the Ising apparatus produces a rather large spectrum of novel insights of collective social phenomena. In particular several new insights are obtained \cite{s1}.
 \begin{itemize}
\item Striking versus working state

A firm, which is either working normally or on strike, appears to be in either one of two symmetrical ordered phases. Accordingly the amplitude of a strike will be identical to the prevailing amplitude of working just before the strike bursts.

\item Avoiding a strike is cheaper than to put a firm back to work

Having a good effective wage with a $H>0$ guarantees a working state with a production value $M>0$. From a working state with $H>0$,  decreasing the effective wage to $H<0$, either by decreasing the actual paid wave $W$ or by an increase of the expected minimum salary $E$, will have no apparent effect. As long as the value of $H<0$ is within the metastability range the working sate can persist with no strike occurring. Only below the limit of metastability does the strike bursts. 

However while within the metastable working state any minority action, including external one, could precipitate the whole firm on strike. The same action would have no effect in the range $H>0$. It is the nucleation phenomenon, which takes place within the striking phase. It explains why it is cheaper to avoid a strike by increasing $W$ to reach a$H>0$ than to put a firm back to work. in the second case a much larger $H$ is required since it has to be beyond the metastability range as seen in Figure (\ref{h4})

\item The disordered state
For $K>K_c$ the firm is in a disordered state characterized by an average individual production with $M=0$. Then the current wage is instrumental to keep the firm working or striking. As soon as $H<0$ the agents strike and as son as $H>0$ they are working. 

\end{itemize}

No precise predictions were drawn from this model up to date. But the field of sociophysics was outlined. Some ethical and epistemological questions were also addressed making the paper \cite{s1} a manifesto for sociophysics in addition to solving a peculiar problem.

\subsection{Consensus versus extremism}

With Moscovici \cite{s2, s3, s4, s5, s6}, a leading social psychologist, we addressed the basic question of polarization and risk taken observed in a series of experiments conducted in experimental psychology. Here, polarization means a consensus on a non-centrist opinion, mainly an extreme one. The unexplained puzzle was to understand why a given group, which has to come up with a decision without time constraint, ends up most of the time at an extreme decision instead of a consensus at a average decision. 

The problem was studied for a two-choice situation using again the Ising ferromagnetic model in an external field with the same formal Hamiltonian given by Eq. (\ref{strike}). But 
the parameters are given different meaning besides for the coupling $J_{i,j}$. However we have introduced an additional local parameter $H_i$  to account for the individual bias agents may have with respect to a series of choices. The bias can vary amplitude and sign from one agent to another making $H_i$ either positive, negative or zero. The 
resulting Hamiltonian is the so called Random Field Ising Model in an external field \cite{s2}
\begin{equation}
{\cal H}=-\sum_{(i,j)} J_{i,j}S_iS_j -\sum_{i=1}^N H_i S_i -H\sum_{i=1}^N S_i ,
\label{deci}
\end{equation}
with $S_i=\pm 1$. In \cite{s2} the variables and parameters are denoted differently with $S_i \rightarrow c_i$, $J_{i,j} \rightarrow I_{i,j}$, $J \rightarrow I$, $H_i \rightarrow S_i$, $H \rightarrow S$, $T\rightarrow D$, $M \rightarrow C/N$ and $e$ is the coordination number. Various extensions and developments of the model were achieved \cite{s7, s8, s10}. A simulation of the model for a small number of agents was performed with Zucker  \cite{s9}.

\subsubsection{Similarities with physical systems and novel counterintuitive social results}

\begin{figure}
\includegraphics[width=\columnwidth]{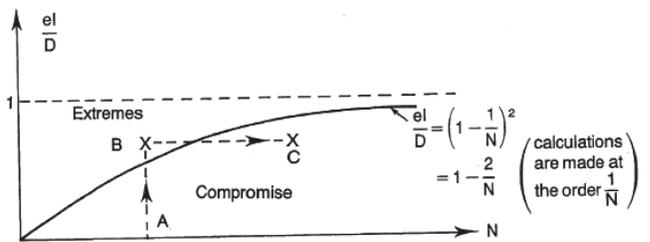}
\caption{Point A represents the compromise choice $C=0$ of a group of N agents without interaction among them ($I=0)$. The vertical arrow to point B shows how beyond some increase in the setting of interactions $I>0$, the same group shifts its choice to a polarized choice $(C\neq 0)$, yet not extreme $(C\neq \pm N)$ since $D>0$. The horizontal arrow to point C indicates that increasing the number of agents of the group within the same external conditions, $I$ and $D$ are unchanged, can turn back the collective choice to a consensus at $C=0$.
}
\label{h5}
\end{figure}

The model used is identical to its statistical counterpart besides two additional points. The first point is a treatment of a finite size system with $N$ agents as shown in Figure
(\ref{h5}). The second point is that while we start with Ising spins $S_i=\pm 1$ once the equilibrium state is reached we shift to continuous spins variables $\tilde S_i$ with $-1\leq {\tilde S_i }\leq 1$.

Although the problem is solved using a mean field treatment, it was given a frame which makes it exact. Moreover the concept of anticipation was materialized via the existence of an order parameter, which is a priori unknown and enters the usual self consistent mean field equation of state.

The associated psychosociology implications have been investigated thoroughly in particular with respect to both opposite phenomena of group polarization and consensus formation. Group polarization leads to the emergence of extremism ($C=\pm N$ while consensus ($C=0$ creates new moderate choices. All choices with $-N\leq C \leq +N$ can be obtained driven by the local field distribution. Several experimental facts which were still without understanding, were given a new coherent explanation. A series of solid consequences regarding the settings of group decision making including juries were outlined. The leader effect was also considered. All these insights and corresponding results are too numerous to be summed here and we refer to the related papers for details \cite{s2, s3, s4, s5, s6, s7, s8, s9, s10}.

\section{Coalitions and fragmentation in a group of countries}
 
Once the Ising ferromagnet has been used to describe social situations it was natural to extend the application to include antiferromagnetic coupling. It was done in a series of models to study the formation of coalitions and the dynamics of fragmentation among a group of entities, where the entities were chosen to be countries \cite{f1, f2, f3, f4, f5}. But the model applies equally to a group of firms or persons or any social bodies.

\subsection{Spontaneous coalition forming and fragmentation}

The model considers a superposition of two kinds of spin glasses, which are a random bond spin glass plus a random site spin glass. A group of $N$ countries is considered with two possible coalitions denoted respectively A and B. A variable $\eta _i$ associated to each country determines its alliance with $\eta _i=+1$  for $i \in A$ and $\eta _i=-1$ when $i \in B$. 

Any pair of countries $(i, j)$ is connected by a bilateral
propensity $G_{i,j}$ to either cooperate $(G_{i,j}>0)$ or to conflict
$(G_{i,j}<0)$. Absence of link is also possible with $(G_{i,j}=0)$. The sign and amplitude of $G_{i,j}$ are quenched variables, which result from the history of the relationship between the two countries $i$ and $j$. 

Propensities $G_{i,j}$ are local and cannot be modified on time scale of years. In parallel
coalitions have been known to exist since long ago, therefore each country has always an a priori natural belonging to one of the two coalitions. We represent this tendency by
a variable $\epsilon _i$ with $\epsilon_i=+1$ when country $i$ would rather join $A$ and $\epsilon _i=-1$ if $i$ would rather join $B$. No natural a priori is denoted by $\epsilon _i=0$.

Monitoring exchanges between a pair of countries $(i,j)$ by a coupling $J_{i,j}>0$, it yields an effective coupling $J_{i,j}\epsilon _i\epsilon j$, which can be either positive or negative depending if countries $i$ and $j$ are cooperating (belongs to the same alliance) or competing (belongs to opposite alliances). An additional propensity $J_{i,j}\epsilon _i\epsilon_j\eta _i\eta _j$ is thus created 

Adding above both propensities gives the total pair propensity
\begin{equation}
p_{i,j}\equiv G_{i,j} +\epsilon _i \epsilon _j J_{i,j} ,
\end{equation}
between countries $i$ and $j$.

An additional variable $\beta_i=\pm 1$ is introduced to account for
benefit from economic and military pressure attached to a given alignment.
It is still
$\beta _i=+1$ in favor of $A$, $\beta _i=-1$ for $B$ and $\beta _i=0$ for
no pressure.
The amplitude of this economical and military interest is measured by a local
positive field $b_i$ which also accounts for the country size and importance.
At this stage, sets $\{\epsilon _i\}$ and $\{\beta _i\}$ are independent local quenched variables.

The respective alignment of two countries $i$ and $j$ is then expressed by
the product $\eta _i\eta _j$. It is $+1$ when $i$ and $j$ belong to the same coalition and $-1$ otherwise. Given all country choices to join either one of the coalitions, a global measure of conflict is thus given by
\begin{equation}
H=-\frac{1}{2}\sum_{i>j}^N\{G_{i,j} +\epsilon _i \epsilon _j J_{ij}\}\eta
_i\eta _j
-\sum_{i}^N \beta _ib_i\eta _i ,
\label{coa}
\end{equation}
where $N$ is the number of countries \cite{f1, f2, f3, f5}. 
 
\subsection{From Ising to Potts variables}

Most coalition settings end up with two competing large alliances, but that is not always the case \cite{f4}. Sometimes more than 2 simultaneous alliances, though not much more are exhibited by the corresponding group of actors. One example is found with computer operating systems where the three different competing systems Windows, Mac OS and  Linux $\backslash$ Unix are used.

Accordingly we extend with Florian \cite{f4} the previous bimodal approach by allowing for multimodal
coalitions. It is achieved substituting Potts variables to the Ising ones. To include all possible cases, the number of Pots states is taken equal to the number $N$ of actors. The number $q$ of actual coalitions is thus turned into an internal degree of freedom, which can vary from (1) to $N$. This extension allows quite naturally for the possibility of neutrality by setting to zero ($\eta _{i}=0$) one of the possible $N$ discrete values. The associated Hamiltonian is
\begin{equation}
{\cal H}=-\sum _{i>j}^{n}J_{ij} \delta (\eta _{i}, \eta
_{j})[1-\delta(\eta _{i}\eta _{j},0)] \ ,
\end{equation}
where the last factor accounts for the case when both $i$ and $j$
are neutral.

While investigating the case of ex-Yugoslavia a new expression for the interaction was suggested with
\begin{equation}
J_{ij}=\sum _{k, l}^8 q_{ik} q_{jl} w_{kl} \ ,
\end{equation}
where $q_{ik}$ represents the percentage of ethnic group $k$ in
entity $i$ and $w_{kl}$ represents the pairwise propensity between
ethnic groups $k$ and $l$ \cite{f4}.

For $k=l$, $w_{kk}=+1$. For $k\neq l$, the $w_{kl}$'s are computed
as the sum of 2 terms. One stands for religion and the other for
language: $w_{kl}=\omega _{religion}(k, l)+\omega _{language} (k,
l)$. For more details we refer to \cite{f1, f2, f3, f4, f5}.

\subsection{Similarities with physical systems and novel counterintuitive social results}

Our first coalition model borrows from statistical physics the random bond spin glass model, the Mattis random site spin glass model, and the random field model. However the novelty was to combine these three models simultaneously in the unique Hamiltonian given by Eq. (\ref{coa}).

Several international situations including the stability of the cold war period and the Eastern Europe instabilities, which followed the auto dissolution of the Warsaw pact, were discussed and given a new surprising insight. Several counterintuitive hints on how to determine some specific international policies were elaborated. For a detailed presentation we refer to  \cite{f1, f2, f3, f4, f5}.

The second version \cite{f4} introduced a novel definition of the pair interaction, which does allow for a direct evaluation using real data.  Several interesting results were found with respect to the second world war and the fragmentation of ex-Yugoslavia. The case of Kosovo was also given some light.


\section{Global versus local terrorism}

After 2001 September 11 with the attack on the US, terrorism has turned to a worldwide permanent threat. We suggest to apply percolation theory to address the problem of the range of possible destruction from a terrorist group \cite{t1, t3, t4, t5}. Instead of studying the terrorist groups themselves we focus the investigation on the what we called the terrorism passive supporters. They are people, who are sympathetic to the terrorism cause, but without any active involvement. They just don't oppose a terrorist move in case they could. Most of them are always dormant and do not need to identify themselves. They are unknown. Only their proportion can be roughly estimated using polls and elections where associated political groups are running.

We start from the fact that terrorism has existed long time ago, and that until
recently it was restricted to very localized geographical areas like for well-known cases of European terrorism in Corsica, Northern Ireland and Euskadi. They are  local terrorism. In contrast, the 2001 September 11 attack on the USA, as well the following attacks on Spain and England,  has revealed a worldwide range of action of a terrorist network. It is a global terrorism. 

\begin{figure}
\includegraphics[width=\columnwidth]{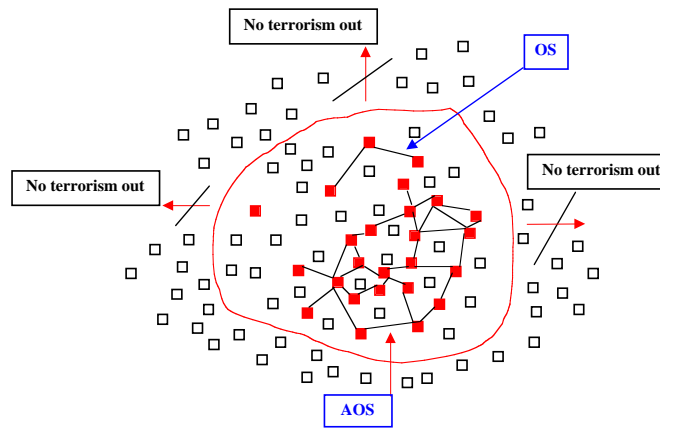}
\caption{Schematic representation of a local terrorism with an Active Open Space, which percolates at the level of 
the island but with no possible extension beyond. Active means that the terrorist base is on it. Black squares are passive supporters while
while ones are not.}
\label{t1}
\end{figure}

Our percolation approach embeds both local and global terrorism within a unique frame. 
According they are two phases of a percolation transition in the assembly of randomly distributed passive supporters to the terrorist cause. The local terrorism corresponds to the disordered phase with only finite sized clusters of connected people. In that case the density $p$ of passive supporters is below the percolation threshold $p_c$ as seen in  Figure (\ref{t1}). The long-range terrorism corresponds to the case $p>p_c$ where the system is ordered with the
existence of an infinite percolating cluster of passive supporters.
There, global terrorism becomes spontaneously achievable. This range of destruction
property is independent of the terrorist net itself.

Within the above frame, the 2001 September 11 attack reveals the existence of
a world percolation of passive supporters. Consequently, the reduction of
the world density of passive supporters below the percolation threshold becomes
the major strategic goal of an efficient fight against this international
terrorism. However, even a few percent reduction of the world passive
supporter density would require neutralizing millions of people, either
physically or ideologically, making both options non-ethics and totally
unpractical within reasonable action. 

At this stage, the conclusion is very pessimistic with no solution to dismiss the current world level of terrorism threat. The lack of solution comes from the fact that $p_c$ being fixed by the ground topology, which is also fixed, to complete
the condition $p<p_c$, where the long range terrorism is defeated, requires to reduce the density $p$ of the passive supporters. But as stated above, it is simultaneously impossible, unacceptable and inefficient to neutralize millions of people.

However, we made with Mauger  the hypothesis that the various independent-fighting goals set by a terrorist group can be represented by a set of independent variables, which in turn
extends the geographic space onto a higher dimensionality virtual 
social space \cite{t2}. 

Only passive supporters populate this space according to their degree of identification. The associated percolation occurs within this virtual social space. Once it happens, the virtual percolating cluster is found to create additional ground pair links on earth surface, which are found to extend beyond the original nearest neighbor distance. The virtual social space dimension monitors the range of these additional ground connections. On this basis the percolation threshold $p_c$  can be modified to reach $p<p_c$ by decreasing the social space dimension, leaving the density $p$ unchanged.

A new strategic scheme to suppress the passive supporter percolation without dealing with the passive supporters themselves can thus be apprehended. Since acting on the population is shown to be useless, the operative goal is to increase the terrorism percolation threshold by reducing the dimension of its social space, which includes both the ground earth surface and all various independent flags displayed by the terrorist group \cite{t2, t5, t6}. The model applies to a large spectrum of clandestine activities including guerilla warfare as well as tax evasion, corruption, illegal gambling, illegal prostitution and black markets \cite{t3}.

\subsection{Similarities with physical systems and novel counterintuitive social results}

From the above analysis it appears that the novelty lies on the way to produce a transition from a percolation condition $p>p_c$ to a non-percolating one $p>p_c$. While in physics $p_c$ is fixed and $p$ is changed to a new lower value $p'$, in our model $p$ is kept fixed and $p_c$ is changed to a higher value $p_c'$. 

The numerous different consequences driven from this approach to fight terrorism are also counterintuitive and may open new ways to implement the political fight against global terrorism.

\begin{figure}[t]
\includegraphics[width=.6\textwidth]{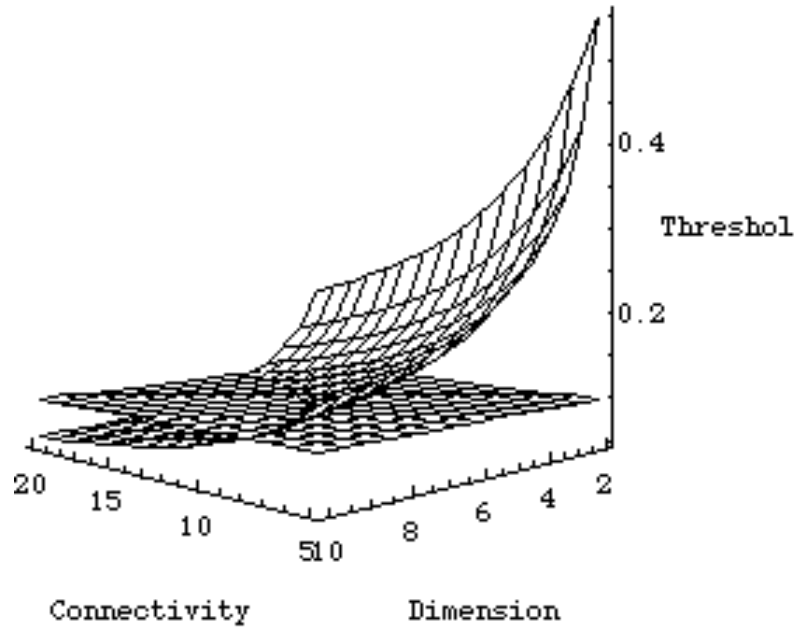}
\caption{Representation of the  Galam-Mauger universal formula for percolation thresholds as function of connectivity and dimension. The formula writes $p_c
= a[(d-1) (q-1)]^{-b}$ where $d$ is dimension, $q$ connectivity, $a= 1.2868$ and $b = 0.6160$.}
\label{t2}
\end{figure}

\begin{figure}
\includegraphics[width=\columnwidth]{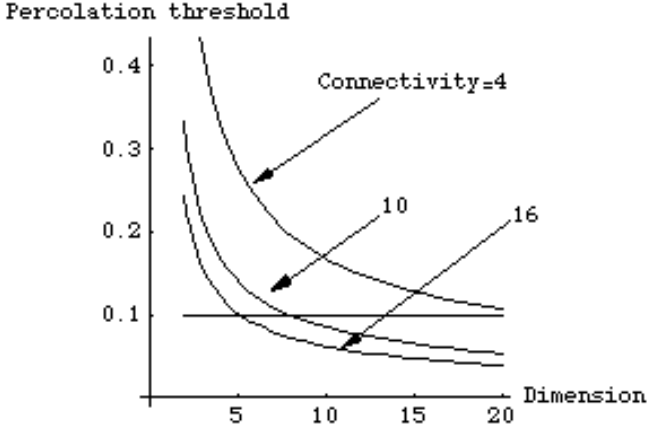}
\caption{Representation of the universal formula Galam-Mauger for fixed
connectivity as function of dimension. It is seen that the threshold values drop with
dimension}
\label{t3}
\end{figure}

The instrumental tool to above approach is the  universal Galam-Mauger formula
for all percolation thresholds  \cite{perco}, 
\begin{equation}
p_c = a[(d-1)(q-1)]^{-b}\ ,
\end{equation}
where $d$ is dimension, $q$ the connectivity, $a= 1.2868$ and $b = 0.6160$. It yields within often excellent accuracy most known thresholds. The formula is shown in three dimensions in Figure (\ref{t2}) and in two dimensions in Figure (\ref{t3}).

\section{Opinions dynamics}

Opinion dynamics has become a main stream of research in sociophysics. The importance of understanding the phenomenon is of a crucial importance in modern democratic societies. Our approach to tackle the question relies on a few simple assumptions, which in turn provide a series of astonishing and powerful results \cite{o1, o2, o3, o4, o5, o6, o7, o8, o9, o10, o11, o12, o13,  o14, o15, o16}.

We consider a population with $N$ agents facing a public debate. It may be prior to a vote, to address a national issue, or to believe a rumor. Two opinions are then competing denoted respectively S and O. For instance supporting the reform or opposed to the reform.

Each agent holds an opinion, either S or O, but it can shift its opinion, driven by two different and decoupled mechanisms. The first one corresponds to all external influences, which apply directly on agents individually, with more or less efficiency. It includes the global  information available,  the private informations some person may have and the mass media. The second mechanism concerns the public debate itself, i.e., the dynamics of individual opinion shifts monitored by the various and repeated discussions among the agents. 

Both level are interpenetrated but here we decoupled them to study specifically the laws governing the internal dynamics. We thus focus on the study of the second mechanism, the first one being taken into account in the values of the initial proportions $p_{S,t}$ and $p_{O,t}\equiv 1-p_{S,t}$ of support for opinions S and O at a time $t$ prior to the starting of the public debate.

The dynamics is then operated via a series of repeated single steps. At each step all agents are distributed randomly among small groups of a few agents, whose size $r$ may vary with $r=1, 2, ...L$. Then within each group all agents adopt the same opinion according to some local rule. A majority-rule is used whenever a local majority exists. Some common belief ``inertia principle" is applied at a tie. There, opinion O is adopted with probability $k$ and opinion S with  probability $(1-k)$, where $k$ accounts for the collective bias produced by the common believes of the group members. After one step, the respective proportions $p_{S,t}$ and $p_{O,t}$ are updated to new values $p_{S,t+1})$ and $p_{O,t+1}$. For even sizes,
\begin{equation}
p_{S,t+1}= \sum_{m=\frac{r}{2}+1}^{r}  {r \choose m} p_{S,t}^m  \{1-p_{S,t}\}^{r-m} +(1-k)p_{S,t}^\frac{r}{2}  \{1-p_{S,t}\}^\frac{r}{2} \ ,
\label{pr-even} 
\end{equation}
where $ {r \choose m}\equiv \frac{r !}{m ! (r-m) !}$ is a binomial coefficient, and for odd sizes
\begin{equation}
p_{S,t+1}= \sum_{m=\frac{r+1}{2}}^{r}  {r \choose m} p_{S,t}^m  \{1-p_{S,t}\}^{r-m} \ .
\label{pr-odd} 
\end{equation}

Steps are repeated $n$ times with corresponding $p_{S,t}, p_{S,t+1}, p_{S,t+2}, ...p_{S,t+n-1}=p_{S,t+n}$ for which an equilibrium state is obtained. The associated dynamics is fast and leads to a total polarization along either one of the two competing states S and O. The direction of the opinion flow is determined by an unstable separator at some critical density $p_{c,r}$ of agents  supporting the S opinion. 

\subsection{The local majority-rule and the existence of doubt}

\begin{figure}
\includegraphics[width=.5\textwidth]{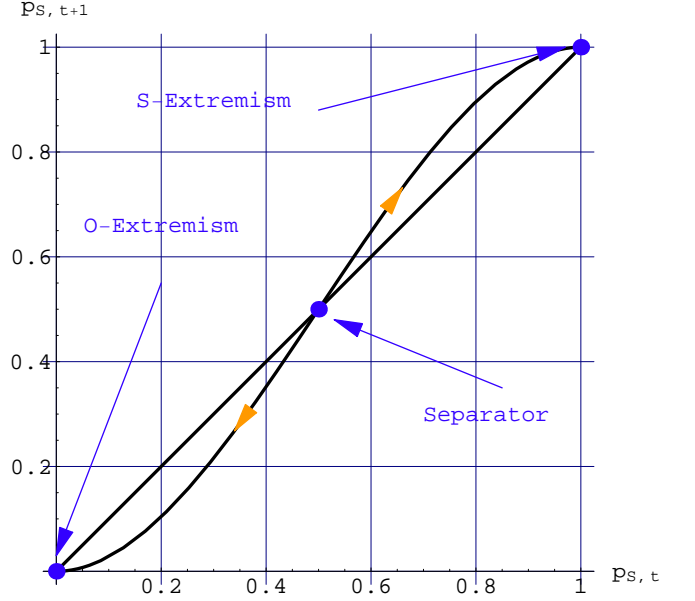}\hfill
\includegraphics[width=.5\textwidth]{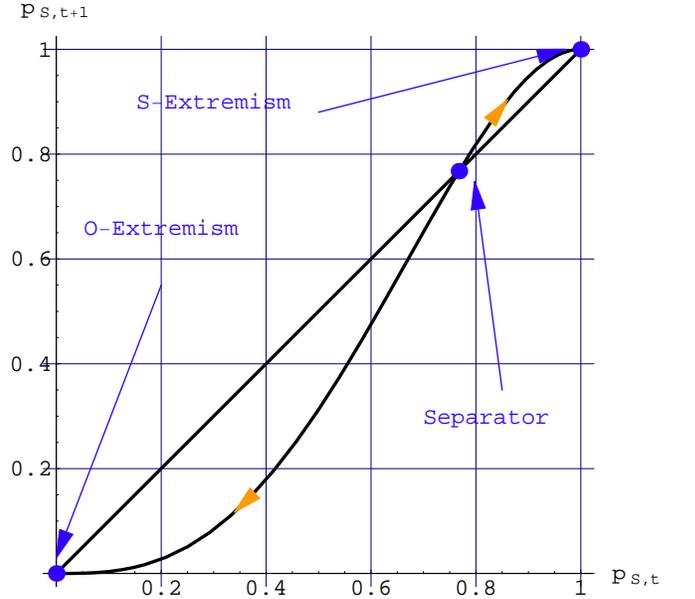}
\caption{On the left side $p_{S,t+1}$ as a function of $p_{S,t+1}$ fo the odd size $r=3$. The right side corresponds to $r=4$ and $k=0$.}
\label{c1-2}
\end{figure}

In the case of odd size groups, $p_{c,r}=1/2$ as shown in Figure \ref{c1-2}. By contrast even sizes make $p_{c}\neq 1/2$. The corresponding 
asymmetry in the dynamics of respectively opinion S and O arises from the existing
of local collective doubts at a tie. The  value of $p_{c,r}$ is a function of $r$ and $k$ \cite{o10}. Figure \ref{c1-2} has $k=0$.

For groups of size $r=4$ with $k=1$ the unstable separator may then be simultaneously 
at a value of $23\%$ for O and at $77\%$ for S. As a function of $k$ we have
\begin{equation}
p_{c,4}=\frac{(6k-5)+\sqrt{13-36k+36k^2}}{6(2k-1)} \ ,
\end{equation}
except at $k=1/2$ where $p_{c,4}=1/2$. 

Considering a distribution of group sizes gets the threshold to cover the all range $0\leq p_c \leq 1$ yielding a very rich and complex phase diagram. 

\subsection{The reshuffling effect and rare event nucleation}

The use of iterated probabilities in the analytical treatment implies a reshuffling of agents between two successive update. This reshuffling effect was investigated using a cellular automata simulation with Chopard, Masselot, and Droz \cite{o1}. It allows to discover the occurrence of very rare events, which can, under some specific conditions, nucleate and invade the all system \cite{o2,o4}. An application to cancer tumor growth was made performing numerical simulations with Radomski \cite{oc1}.

\subsection{Extension to 3 competing opinions and size combinations}

With S. Gelke and L. Peliti we investigated the flow diagram of opinion dynamics for three competing opinions A, B and C in the case of update local groups of size 3 \cite{o11}. A local majority-rule is applied when possible. The case of a tie (A, B, C) yields three different updates, which are (A, A, A) with a probability $\alpha$, (B, B, B) with a probability $\beta$, and (C, C, C) with a probability $(1-\alpha-\beta)$. 

The associated flow diagram is two dimensional and exhibits a very rich variety of highly non linear behaviors.  Several fixed points are involved.

The model was also extend to consider a distribution of local groups whose sizes may vary from $r=1$ up to $r=L$ where $L$ is some integer usually smaller than 10 \cite{o3}.

\subsection{Heterogeneous beliefs, contrarian and inflexible effects}

In real society not everyone discusses with every one, people are divided within subgroups which shares different collective believes \cite{o10}. Therefore we extended our model to include heterogeneous beliefs with different collective opinions, which add eventually to yield the global opinion of the society.

In addition to above heterogeneity we applied with Pajot, percolation theory to address the problem of coexistence of opposite collective opinions, like for instance with the feeling of safeness, within the same social frame. The possibility of superposition of simultaneous percolation from two different species was analyzed and shown to provide an explanation to such a paradoxical  social phenomenon \cite{oo1}.

Coming back to the our opinion dynamics model we accounted for the fact that not everybody is to be convinced by a majority of arguments. On the contrary sometimes, agents behave oppositely making  contrarian choices \cite{o6, o15}. A contrarian is someone who deliberately decides to oppose the prevailing choice of the majority around it whatever is that choice. 

For low density of contrarians, below some threshold, the separator is left unchanged at fifty percent while the two pure attractors, total A or total B, are shifted towards mixed phases attractors. A majority of A (B) coexists with a minority of B (A). However above this threshold the dynamics is reversed with only one single attractor at fifty percent. the equilibrium state becomes a perfect equality between the two competing opinions A and B as seen ifn Figure Öref{c3-4}

\begin{figure}
\includegraphics[width=.5\textwidth]{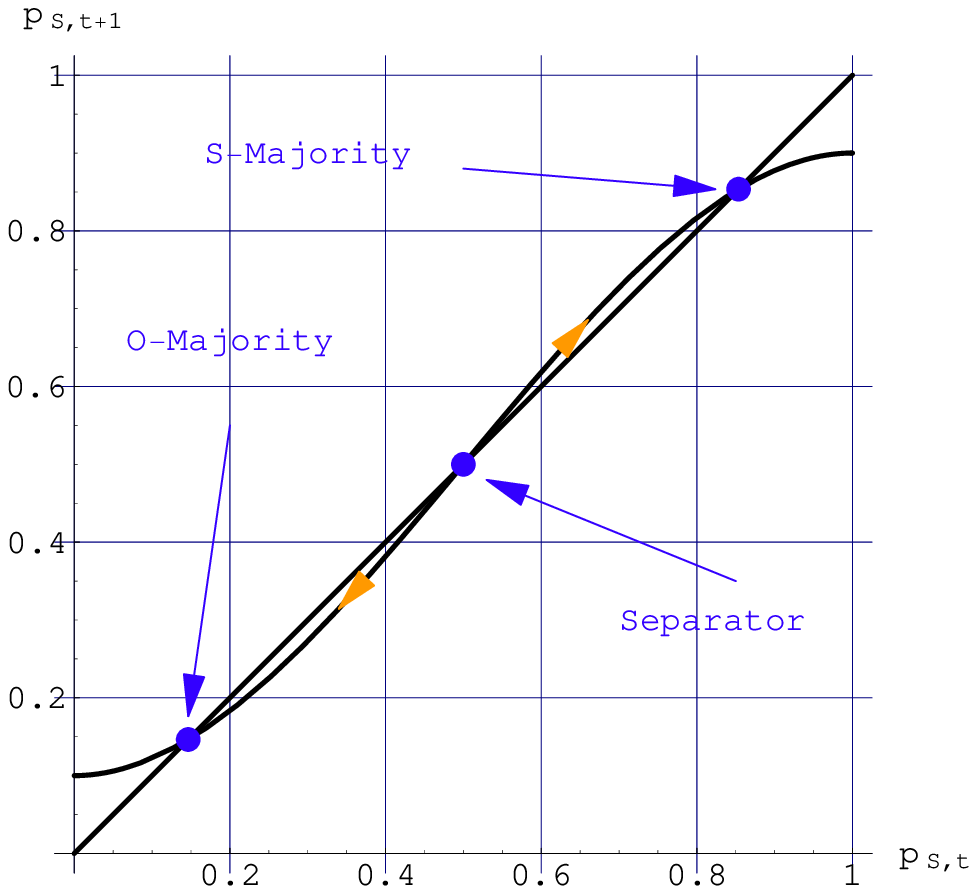}\hfill
\includegraphics[width=.5\textwidth]{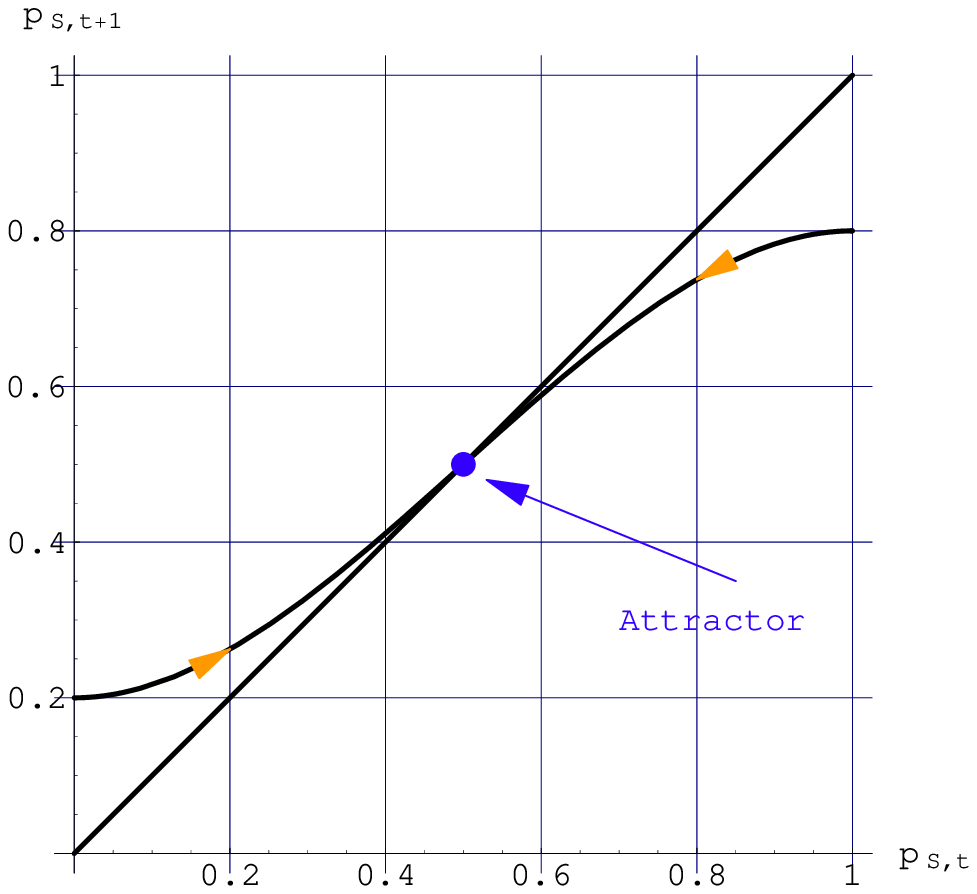}
\caption{On the left side $p_{S,t+1}$ as a function of $p_{S,t+1}$ fo the odd size $r=3$ with a density of $10\%$ of contrarians. The right side corresponds to a density of $20\%$ of contrarians..}
\label{c3-4}
\end{figure}

With Borghesi we extend the contrarian behavior to an opposition to the global choice given by polls \cite{o13}. The effect is similar to the previous one but now a chaotic behavior is obtained around fifty percent.

Another feature of human character is the inflexible attitude. With Jacobs we introduced in the opinion dynamics agents who stick to their opinion whatever argument is given to them, they are inflexible \cite{o16}. The effect is similar to the contarian effect, but an asymmetry is found between the two opinions A and B. It depends on the the difference in the respective proportions of inflexibles. The single attractor is no longer at fifty percent making certain the victory of the opinion, which has more inflexibles.

\subsection{Similarities with physical systems and other sociophysics models}

\begin{figure}
\includegraphics[width=\columnwidth]{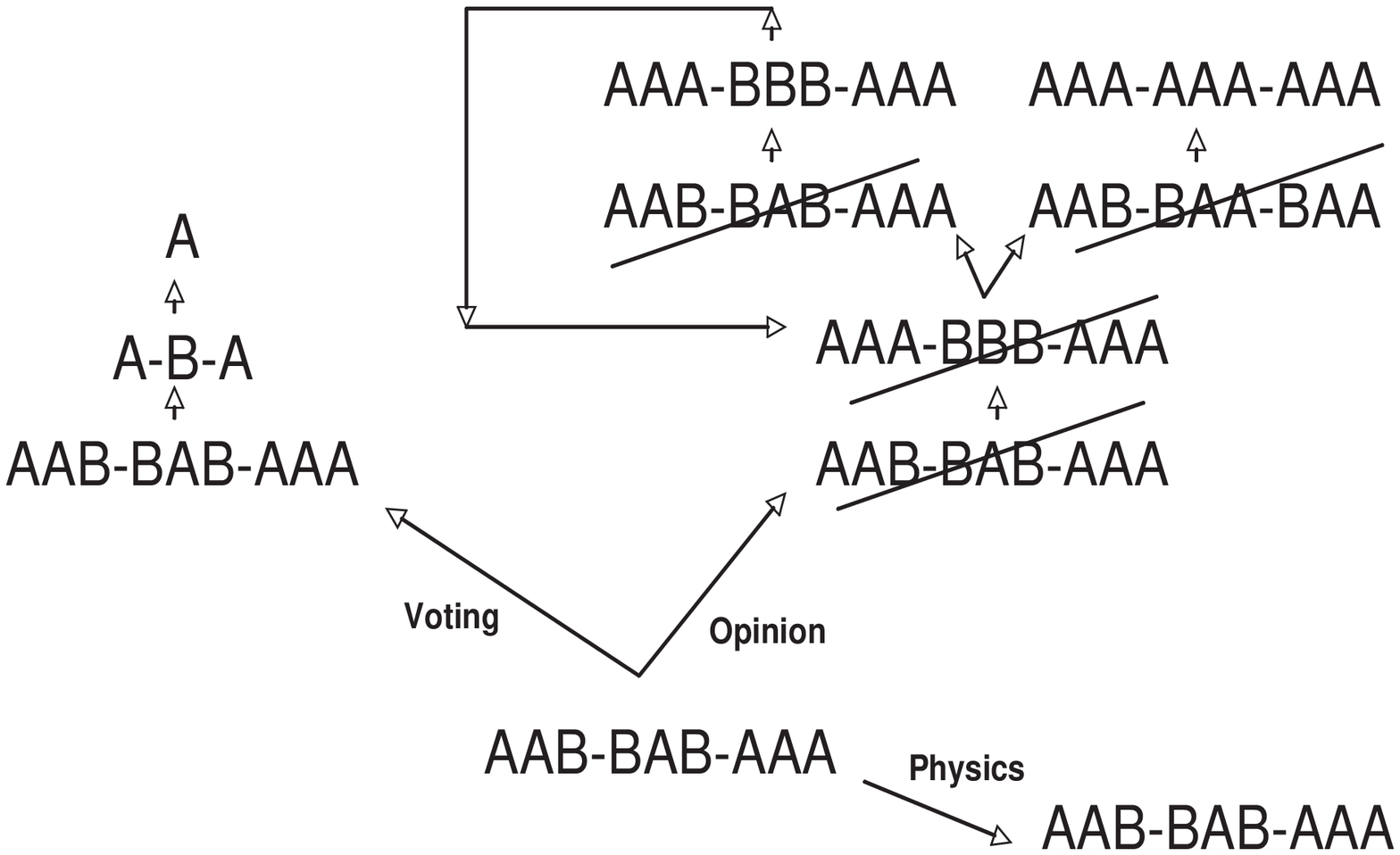}
\caption{On the left side $p_{S,t+1}$ as a function of $p_{S,t+1}$ fo the odd size $r=3$ with a density of $10\%$ of contrarians. The right side corresponds to a density of $20\%$ of contrarians..}
\label{ga}
\end{figure}

It is worth to stress that when an update occurs it is applied to the same and full  population of agents. Agents do shift their opinion eventually. It is different from our voting model where groups of agents elect representatives as shown in Figure (\ref{ga}). 

For instance in the opinion dynamics model the configuration $(AAB)$, $(BAB)$, $(AAA)$ yields  $(AAA)$, $(BBB)$, $(AAA)$ after one update. At opposite in the voting model the same configuration $(AAB)$, $(BAB)$, $(AAA)$ yields the addition of the group $(ABA)$ above it. In the first case, the agents are the same but their respective opinion may have changed. In the second case, the agents do not shift opinion but additional agents are added according to their respective distribution within the local groups. 

Performing one additional update in the voting model add a second level with one agent $(A)$, the president while in the opinion model the update requires first a reshuffling. In our case two kind of configurations $(AAB)$, $(BAB)$, $(AAA)$ or $(AAB)$, $(BAB)$, $(AAA)$ are possible. Then the update is implemented leading respectively to $(AAA)$, $(BBB)$, $(AAA)$ and $(AAA)$, $(AAA)$, $(AAA)$. Some agents have again modified their opinion. 

At this stage no more update can be performed in the voting model, unless the initial assembly of 9 agents is increased to 27 agents. In the opinion model, another update can be performed from the configuration $(AAA)$, $(BBB)$, $(AAA)$.

Our illustration demonstrates what should be always kept in mind, that the same mathematical equation, here a local majority rule, can create two totally different realities. 
 Although the mathematical equations between our two model of voting and opinion are identical, the content, the meaning, the implementation and the result are totally different. 
 
The same remark and conclusion applies with respect to the connection to statistical physics. Applying a real space renormalization group transformation to our initial configuration is meaningless due to the finite size of the sample. However it is possible taking a larger sample. However the use of renormalization group techniques will allow to extract some properties of the sample, but will not modify the sample at all. 

As seen in Figure (\ref{ga}) the same equation may be used for three different models and contents. At opposite, different equations may indeed yield the same model and content. Several different models of opinion dynamics were thus shown to be equivalent  \cite{o12}.

\subsection{Novel counterintuitive social results}

Our model produced a large spectrum of results and several predictions were formulated. Most of them were general without precise details, like the prediction of an increase of occurrence of voting at fifty-fifty in democratic countries. However one precise prediction, the victory of the no to the 2005 French referendum on European Constitution, was stated several months ahead of the event, against all predictions and expectations, and was eventually validated. For more details we refer to the corresponding papers  \cite{op1, op2, op3, op4, op5, op6, op7, op8}. 

The model was also applied respectively  to rumor \cite{o5}, and with Vignes to fashion phenomena \cite{o9}. 

\section{Conclusion}

We present a sketchy overview of five families of Galam and Galam et al models developed to tackle different social and political problems. The aim was to gather for the first time all these works together to allow an easy connection to these works.  A personal testimony about sociophysics can be found in  \cite{e1}. It is also worth to notice that  sociophysics has been able to also produce new results in statistical physics: with Sousa and  Malarz, performing Monte Carlo simulations we discovered new results associated with the two-dimensional reshuffled Ising ferromagnet \cite{e2}.

At this stage, the new challenge of sociophysics is to prove itself it can become a predictive science with well established elementary rules of social and political behaviors. The task is hard but may be achievable.

\section*{ Acknowledgment }

I would like to wish a happy and active retirement to my ``numerical competitor" Prof. Dr. D. Stauffer.


\end{document}